\documentclass[10pt,twocolumn]{article}

\usepackage[left=0.75in,
	top=1in,
	right=0.75in,
	bottom=1in]{geometry}
\usepackage{algorithm}
\usepackage{algpseudocode}
\usepackage{amsmath}
\usepackage{cite}
\usepackage{enumitem}
\usepackage{graphicx}
\usepackage[colorlinks,
	citecolor=blue,
	urlcolor=blue, 
	linkcolor=blue]{hyperref}
\usepackage{mathtools}
\usepackage{times}
\usepackage{xfrac}

\begin{document}

\title{\textbf{A Tutorial on Computing $t$-Closeness}}

\author{Richard Dosselmann, Mehdi Sadeqi, and Howard J. Hamilton\\ \\
	\small Department of Computer Science, University of Regina\\
	\small 3737 Wascana Parkway,
	Regina, Saskatchewan, Canada~S4S 0A2\\
	\small +306-585-4633 (telephone), +306-585-4745 (fax)\\
	\small dosselmr@cs.uregina.ca,
	sadeqi2m@cs.uregina.ca
	and Howard.Hamilton@uregina.ca}

\date{}
\maketitle

\begin{abstract}
This paper presents a tutorial of the computation of $t$-closeness. An established model in the domain of privacy preserving data publishing, $t$-closeness is a measure of the earth mover's distance between two distributions of an anonymized database table. This tutorial includes three examples that showcase the full computation of $t$-closeness in terms of both numerical and categorical attributes. Calculations are carried out using the definition of the earth mover's distance and weighted order distance. This paper includes detailed explanations and calculations not found elsewhere in the literature. An efficient algorithm to calculate the $t$-closeness of a table is also presented.
\end{abstract}

\vspace{10pt}

\noindent\textbf{Keywords:} $t$-closeness, earth mover's distance, privacy preserving data publishing, anonymization

\section{Introduction}
\label{SEC_Introduction}

Database tables are routinely published online. In many instances, this can be readily done without any adverse effects. At times though tables contain sensitive information that should not be disclosed to the public. Attributes with values that directly identify a person or entity, such as name, student ID number, or product code, are \textit{explicit identifers} and are not generally made public \cite{REF_tClose, REF_tClose2}. A set of attributes that together (and indirectly) identifies a person or entity, such as occupation, nationality, and neighborhood, is a \textit{quasi-identifier} (QI) \cite{REF_tClose, REF_tClose2}. Finally, an attribute with values that should not be linked to a person or entity for privacy reasons, such as an employee's salary, a patient's medical condition, or a student's grade, is a \textit{sensitive} attribute \cite{REF_LDiv}. Before publishing a table that contains sensitive information, certain values are either removed or obfuscated, a process known as \textit{anonymization} \cite{REF_DBAnon}. This is done in order to prevent a subsequent user from linking a person or entity to a sensitive value in that table. The study and design of anonymization algorithms is a major topic in the domain of privacy preserving data publishing (PPDP) \cite{REF_PPDP}.

As an example, original and anonymized tables relating to incidents that have occurred at different homes in a city are presented in Tables \ref{TAB_Incidents} and \ref{TAB_IncidentsAnon}, respectively. The explicit identifier \textit{Address} of Table \ref{TAB_Incidents} has been completely removed from the anonymized data of Table \ref{TAB_IncidentsAnon}, while the quasi-identifier \textit{Zone} and sensitive (categorical) attribute \textit{Incident} have been preserved. The individual records of Table \ref{TAB_IncidentsAnon} are further grouped into \textit{equivalence classes}, subsets of (indistinguishable) records having the same QI values \cite{REF_tClose, REF_EqClass}. There are four such equivalence classes in Table \ref{TAB_IncidentsAnon}. Although ``anonymized'', Table \ref{TAB_IncidentsAnon} still reveals much to an observer. A prospective home buyer would know for instance that homes in zone 2C experience frequent power outrages, a finding that might discourage that buyer from purchasing a house in that area. An insurance provider looking at this data might be wary of insuring customers in zone 4F, given the serious events that have taken place in that zone. By comparison, potential buyers or insurers would be more open to the homes and occupants of zone 3B, an area with comparatively minor problems.
\begin{table}[h]
\centering
\begin{tabular}{lll}
\hline
	Address & Zone & Incident\\
\hline
	4984 Apple St. & 2C & power outage\\
	4810 Cherry Ave. & 2C & power outage\\
	4075 Grape Blvd. & 2C & power outage\\
	459 Red Cres. & 4F & theft\\
	754 Blue Blvd. & 4F & fire\\
	541 Yellow Lane & 4F & fatal accident\\
	622 Green Ave. & 4F & fire\\
	14002 Square St. & 9A & sidewalk repair\\
	11411 Circle Cres. & 9A & power outage\\
	2032 Rain St. & 3B & pest control\\
	2055 Snow Ave. & 3B & power outage\\
	2091 Cloudy Ave. & 3B & sidewalk repair\\
	2002 Fog Blvd. & 3B & tree replanting\\
	2040 Sunshine St. & 3B & sidewalk repair\\
\hline
\end{tabular}
\caption{Incidents occurring at homes in a city}
\label{TAB_Incidents}
\end{table}
\begin{table}[h]
\centering
\begin{tabular}{lll}
\hline
	Address & Zone & Incident\\
\hline
	$$* & 2C & power outage\\
	$$* & 2C & power outage\\
	$$* & 2C & power outage\\
\hline
	$$* & 4F & theft\\
	$$* & 4F & fire\\
	$$* & 4F & fatal accident\\
	$$* & 4F & fire\\
\hline
	$$* & 9A & sidewalk repair\\
	$$* & 9A & power outage\\
\hline
	$$* & 3B & pest control\\
	$$* & 3B & power outage\\
	$$* & 3B & sidewalk repair\\
	$$* & 3B & tree replanting\\
	$$* & 3B & sidewalk repair\\
\hline
\end{tabular}
\caption{Anonymized incidents occurring at homes in a city}
\label{TAB_IncidentsAnon}
\end{table}

As the example of Tables \ref{TAB_Incidents} and \ref{TAB_IncidentsAnon} demonstrates, anonymization often fails to fully mask all of the distinguishing content of a table. This motivates one to (formally) quantify the level of privacy of an anonymized table. A number of models to do so have been proposed. One of the earliest, \textit{$k$-anonymity} is the minimum number of (QI) indistinguishable records in a table, where higher values of $k$ are generally representative of a more secure table \cite{REF_kAnon}. In the first equivalence class of Table \ref{TAB_IncidentsAnon}, there are three indistinguishable records, whereas there are four, two, and five such records, respectively, in the three remaining equivalence classes. As a result, $k =\min(3,4,2,5) = 2$, meaning that Table \ref{TAB_IncidentsAnon} is $2$-anonymous. Suppose that a user is interested in learning the value of the attribute \textit{Incident} of a home in zone 2C. Though the user is not able to determine precisely which record in the first equivalence class of Table \ref{TAB_IncidentsAnon} is associated with that specific home, the user is still able to infer that if it experienced an incident, it experienced a power outage, because all homes in that equivalence class have been affected by power outages. Thus, $k$-anonymity does not always accurately represent the actual level of privacy of a table.

Machanavajjhala et al. propose \textit{$\ell$-diversity} as a way of countering the faults of $k$-anonymity \cite{REF_LDiv}. In an $\ell$-diverse table, there are at least $\ell$ ``well represented'' values of a sensitive attribute in each equivalence class. In the first equivalence class of Table \ref{TAB_IncidentsAnon}, there is only one distinct value of the sensitive attribute \textit{Incident}, namely ``power outage'', whereas there are three, two, and four values, respectively, in the remaining equivalence classes. As a result, Table \ref{TAB_IncidentsAnon} is $\min(1,3,2,4)\textrm{-diverse} = 1$-diverse. Consider the second equivalence class of Table \ref{TAB_IncidentsAnon}, pertaining to homes in zone 4F. Even though there are three distinct values of the sensitive attribute, namely ``theft'', ``fire'', and ``fatal accident'', all three pertain to serious events, allowing a user interested in a home in that area to conclude that an associated incident, whatever it might have been, was rather serious. In the end, this means that while $\ell$-diversity considers the number of distinct values of a sensitive attribute, it does not take into account the inherent meaning of those values.

Li et al. introduce $t$-closeness so as to overcome the problems that affect both $k$-anonymity and $\ell$-diversity \cite{REF_tClose, REF_tClose2}. Conceptually, $t$-closeness is the maximum of the distances between the distribution of values of a sensitive attribute of the equivalence classes (of a table) and that of the (entire) table. The intuition is that individual equivalence classes of a table that are similar to that table do not generally reveal any more information than the table itself. Formally, ``an equivalence class is said to have $t$-closeness if the distance between the distribution of a sensitive attribute in this class and the distribution of the attribute in the whole table is no more than a threshold $t$. A table is said to have $t$-closeness if all equivalence classes have $t$-closeness'' \cite{REF_tClose}. At this point, no universal threshold for $t$ appears to exist. The computation of the $t$-closeness of Table \ref{TAB_IncidentsAnon} is deferred to Section \ref{SUBSEC_CatAtt}.

This paper provides a comprehensive tutorial of the calculation of the $t$-closeness model in situations involving either a numerical or categorical attribute. Situations not relating to numerical and categorical attributes are explored in \cite{REF_tClose, REF_tClose2}. The examples presented in this paper incorporate explanations and calculations that go beyond those of the existing literature \cite{REF_tClose, REF_tClose2, REF_CattClose, REF_DBAnon, REF_FuzzyPartitions, REF_KLDivSemanticPrivacy, REF_Health, REF_Microaggregation, REF_SABRE, REF_SurveyThesis, REF_tComplexity}. Additionally, a more efficient algorithm to compute $t$-closeness in circumstances relating to numerical attributes is shown. This paper will ultimately help others to calculate $t$-closeness, as well as provide them with examples for use in testing.

The following section introduces the mathematics behind the calculation of $t$-closeness. Three examples are then presented in Sections \ref{SEC_ExSalary}, \ref{SEC_ExMeritPts}, and \ref{SEC_ExDisease}. An efficient algorithm is given in Section \ref{SEC_EfficientEMDAlgorithm}, followed by some closing remarks in Section \ref{SEC_Conclusion}.

\section{Earth Mover's Distance}
\label{SEC_EMD}

Formally, $t$-closeness is computed using the one-dimensional earth mover's distance (EMD) \cite{REF_EMD}. Two variants of the EMD are examined in this paper. The first is used in situations involving a numerical attribute, as described in Section \ref{SUBSEC_NumAtt}. The second, described in Section \ref{SUBSEC_CatAtt}, is used in the context of a categorical variable.

\subsection{Numerical Attribute}
\label{SUBSEC_NumAtt}

Conceptually, the EMD, as it relates to a numerical attribute, is the total cost of optimally moving \textit{masses} of earth in a space to \textit{holes} in that same space, thus transforming the distribution of the masses to match that of the holes \cite{REF_EMDFaceRecognition}. In the domain of PPDP, attention is restricted to a one-dimensional space in which successive holes are spaced at a distance of one unit apart. In this context, the probabilities $p_i$ of a distribution $\textbf{P}$ denote the masses, while those of a second distribution $\textbf{Q}$, identified as $q_j$, refer to the holes, where $|\textbf{P}|=|\textbf{Q}|= m$ and $1\leq i, j\leq m$. It is assumed that both $\textbf{P}$ and $\textbf{Q}$ are normalized distributions, that is
\begin{equation}
\sum_{i=1}^m p_i =\sum_{j=1}^m q_j = 1.
\end{equation}

As an example, let $Q'_1 =\{14, 27, 88, 101\}$ and $P'_1 =\{14, 88\}\subseteq Q'_1$ be sets of values, where $P'_1$ is an equivalence class of $Q'_1$. Although numbers, the elements of $P'_1$ and $Q'_1$ are effectively nothing more than labels. What matters in this context is not the particular choice of labels, but instead the underlying probabilities of these elements. Each of the values $14$, $27$, $88$, and $101$ appears once in $Q'_1$, meaning that the four probabilities, or holes, of the associated distribution $\textbf{Q}_1$ are of size $q_1 = q_2 = q_3 = q_4 = 1/|Q'_1| =\sfrac{1}{4}$. Thus, $\textbf{Q}_1 =\{\sfrac{1}{4},\sfrac{1}{4},\sfrac{1}{4},\sfrac{1}{4}\}$. Because each of $14$ and $88$ occurs once in $P'_1$, they are assigned probabilities, or masses, of $p_1 = p_3 =\sfrac{1}{2}$. The two remaining values of $Q'_1$, specifically $27$ and $101$, are not found in $P'_1$. They are accordingly given masses of $p_2 = p_4 = 0$. It follows that the distribution of $P'_1$ is $\textbf{P}_1 =\{\sfrac{1}{2}, 0, \sfrac{1}{2}, 0\}$. Because $P'_1$ has only two elements, $\textbf{P}_1$ is extended using zeros to ensure that it, like $\textbf{Q}_1$, is of length $m = 4$.

Two ways of calculating the EMD in the case of a numerical attribute are given by Li et al. \cite{REF_tClose, REF_tClose2}. The first is based on the definition of the EMD, while the second relates to the underlying transformation given by the EMD. These two ways are empirically shown to be equivalent in each of the examples of this paper.

\subsubsection{Calculation using the Definition}
\label{SUBSUBSEC_EMDDefn}

The first way of calculating the EMD employs the definition \cite{REF_tClose, REF_tClose2}, which is
\begin{equation}
E(\textbf{P},\textbf{Q})
=\frac{1}{m-1}\sum_{i=1}^m
	\left|\sum_{j=1}^i\left(p_j - q_j\right)\right|.
\label{EQ_EMD}
\end{equation}
Equation (\ref{EQ_EMD}) represents the cost of transforming a distribution $\textbf{P}$ into another distribution $\textbf{Q}$. Using Equation (\ref{EQ_EMD}), the EMD, and hence $t$-closeness, of $P'_1$ and $Q'_1$ of the previous section is
\begin{gather}
E(\textbf{P}_1,\textbf{Q}_1)
=\frac{1}{4-1}\left[
	\left|\frac{1}{2}-\frac{1}{4}\right| +\right.
	\nonumber\\
\left.\left|\left(\frac{1}{2}-\frac{1}{4}\right) +
	\left(0-\frac{1}{4}\right)\right| +\right.
	\nonumber\\
\left.\left|\left(\frac{1}{2}-\frac{1}{4}\right) +
	\left(0 -\frac{1}{4}\right) +
	\left(\frac{1}{2}-\frac{1}{4}\right)\right|+\right.
	\nonumber\\
\left.\left|\left(\frac{1}{2}-\frac{1}{4}\right) +
	\left(0 -\frac{1}{4}\right) +
	\left(\frac{1}{2}-\frac{1}{4}\right) +
	\left(0-\frac{1}{4}\right)\right|\right]
	\nonumber\\
\approx 0.1667.
\label{EQ_EMDEx}
\end{gather}

When there is more than one equivalence class in a given table, the $t$-closeness of that table is the maximum of the EMD values of the individual equivalence classes.

\subsubsection{Calculation using the Weighted Ordered Distance}
\label{SUBSUBSEC_EMDWeightedOrderedDist}

The second way of calculating the EMD requires that one first define a notion of the distance that separates the elements of the distributions $\textbf{P}$ and $\textbf{Q}$. Let $P'$ and $Q'$ be totally ordered \cite{REF_TotalOrder} multisets \cite{REF_Multiset} with elements sorted in ascending order by frequency (structures that are hereinafter referred to as \textit{sets}) associated with $\textbf{P}$ and $\textbf{Q}$, respectively. Note that elements do not need to be ordered in the case of a categorical attribute. Multisets are used in place of conventional sets since elements are often repeated, such as in the case of $\{3, 3, 3, 1, 1, 2, 4\}\neq\{3, 1, 2, 4\}$. Moreover, by enforcing a total ordering, one can be sure that for any $x, y, z\in P'$:
\begin{enumerate}
\item $x\leq x$ (reflexive \cite{REF_ReflexiveAntisymmetricTransitive})
\item $x\leq y$ and $y\leq x \Rightarrow x = y$ (antisymmetric
\cite{REF_ReflexiveAntisymmetricTransitive})
\item $x\leq y$ and $y\leq z\Rightarrow x\leq z$ (transitive
\cite{REF_ReflexiveAntisymmetricTransitive})
\item $x\leq y$ or $y\leq x$ (trichotomy \cite{REF_Trichotomy})
\end{enumerate}

With this framework established, the \textit{ordered distance} \cite{REF_tClose, REF_tClose2} between elements $v_i\in P'$ and $v_j\in Q'$ is
\begin{equation}
D(v_i,v_j) =\frac{|i-j|}{m-1}.
\label{EQ_OrderedDist}
\end{equation}
Equation (\ref{EQ_OrderedDist}) does not take into account the mass moved between $v_i$ and $v_j$, only the distance separating these two values. Thus, in this paper, the right-hand side of Equation (\ref{EQ_OrderedDist}) is multiplied by the actual amount $w_{i,j}$ of mass transferred between $v_i$ and $v_j$, resulting in the \textit{weighted ordered distance}, defined as
\begin{equation}
D'(v_i,v_j) = w_{i,j}\cdot\frac{|i-j|}{m-1}.
\label{EQ_WeightedOrderedDist}
\end{equation}
Consequently, the EMD of Equation (\ref{EQ_EMD}) is also given as
\begin{gather}
E(\textbf{P},\textbf{Q}) =\sum_{i=1}^n D'(v_i,v_j),
\label{EQ_EMDWeightedOrderedDist}
\end{gather}
where $n$ is the number of optimal actions needed to transform $\textbf{P}$ into $\textbf{Q}$.

\begin{figure}[h]
\centering
\includegraphics[]{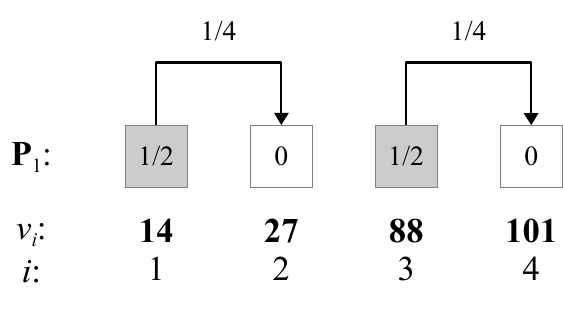}
\caption{Transformation of $\textbf{P}_1$ to $\textbf{Q}_1$}
\label{FIG_ExWeightedOrderedDist}
\end{figure}

Referring again to the example of $P'_1$ and $Q'_1$ of Section \ref{SEC_EMD}, two (optimal) actions are needed to transform $\textbf{P}_1$ into $\textbf{Q}_1$. First, one transfers $w_{1,2}=\sfrac{1}{4}$ from mass $p_1 =\sfrac{1}{2}$ of $v_1 = 14\in P'_1$ from index $i=1$ to hole $v_2 = 27\in Q'_1$ of size $q_2 =\sfrac{1}{4}$ at index $j=2$. Using Equation (\ref{EQ_WeightedOrderedDist}), the cost of doing so is
\begin{equation}
D'(v_1, v_2) =\frac{1}{4}\cdot\frac{|1-2|}{4-1}\approx 0.0833.
\end{equation}
In this example, only one-half of the mass $p_1 =\sfrac{1}{2}$ of $v_1 = 14\in P'_1$ is moved to hole $v_2 = 27\in Q'_1$ of size $q_2 =\sfrac{1}{4}$ because only one-half of the mass of $v_1$ can ``fit'' in hole $v_2$ of size $q_2$ ($\sfrac{p_1}{2} =\sfrac{1}{4}$). In the second step, one moves $w_{3,4}=\sfrac{1}{4}$ from mass $p_3 =\sfrac{1}{2}$ of $v_3 = 88\in P'_1$ from index $i=3$ to hole $v_4 = 101\in Q'_1$ of size $q_4 =\sfrac{1}{4}$ at index $j=4$, at a cost of
\begin{equation}
D'(v_3,v_4)
	=\frac{1}{4}\cdot\frac{|3-4|}{4-1}\approx 0.0833.
\end{equation}
Then, by Equation (\ref{EQ_EMDWeightedOrderedDist}),
$E(\textbf{P}_1,\textbf{Q}_1) = 0.0833 + 0.0833\approx 0.1667$, a result that is equal to the value of Equation (\ref{EQ_EMDEx}). These two actions, illustrated in Figure \ref{FIG_ExWeightedOrderedDist}, effectively transform the distribution $\textbf{P}_1$ of $P'_1$ to that of the global distribution $\textbf{Q}_1$ of $Q'_1$.

It is important to recognize that the method of this section requires that the actions that transform a distribution $\textbf{P}$ into $\textbf{Q}$ be chosen optimally. This means that an arbitrary sequence of actions may not yield the lowest possible cost. Li et al. offer no method of determining the optimal sequence of actions for a given problem \cite{REF_tClose, REF_tClose2}.

\subsection{Categorical Attribute}
\label{SUBSEC_CatAtt}

In the case of a categorical attribute, the EMD is computed using the \textit{variational distance} \cite{REF_tClose, REF_tClose2}, specifically
\begin{equation}
E'(\textbf{P},\textbf{Q})
=\frac{1}{2}\sum_{i=1}^m\left|p_i - q_i\right|,
\label{EQ_EMDVarDist}
\end{equation}
a measure that is equal to one-half of the \textit{Manhattan distance} \cite{REF_ManhattanDist}.

As an example, consider again the scenario put forward in Table \ref{TAB_Incidents}. The (global) set of values of the sensitive attribute \textit{Incident} is given as
\begin{gather*}
Q'_2 =\{\textrm{power~outage},
	\textrm{theft},
	\textrm{fire},
	\textrm{fatal~accident},\\
	\textrm{sidewalk~repair},
	\textrm{pest~control},
	\textrm{tree~replanting}\}.
\end{gather*}
The associated distribution of $Q'_2$ (with the values in the same order as that of $Q'_2$) is equal to $\textbf{Q}_2 =\{\sfrac{5}{14}, \sfrac{1}{14}, \sfrac{2}{14}, \sfrac{1}{14}, \sfrac{3}{14}, \sfrac{1}{14}, \sfrac{1}{14}\}$. The first equivalence class of Table \ref{TAB_IncidentsAnon}, denoted
\begin{gather*}
P'_{2,1} =\{\textrm{power~outage},
	\textrm{power~outage},
	\textrm{power~outage}\},
\end{gather*}
contains three occurrences of the value ``power outage'', yielding a distribution of $\textbf{P}_{2,1} =\{\sfrac{3}{3}, 0, 0, 0, 0, 0, 0\}$. The three remaining equivalance classes are
\begin{gather*}
P'_{2,2} =\{\textrm{theft},
	\textrm{fire},
	\textrm{fatal~accident},
	\textrm{fire}\}
\end{gather*}
($\textbf{P}_{2,2} =\{0, \sfrac{1}{4}, \sfrac{2}{4},
	\sfrac{1}{4}, 0, 0, 0\})$,
\begin{gather*}
P'_{2,3}
	=\{\textrm{sidewalk~repair}, \textrm{power~outage}\}
\end{gather*}
($\textbf{P}_{2,3} =\{\sfrac{1}{2}, 0, 0, 0, \sfrac{1}{2}, 0, 0\})$,
and
\begin{gather*}
P'_{2,4} =\{\textrm{pest~control},
	\textrm{power~outage},
	\textrm{sidewalk repair},\\
	\textrm{tree~replanting},
	\textrm{sidewalk~repair}\}
\end{gather*}
($\textbf{P}_{2,4} =\{\sfrac{1}{5}, 0, 0, 0,
\sfrac{2}{5}, \sfrac{1}{5}, \sfrac{1}{5}\}$),
respectively.

Using Equation (\ref{EQ_EMDVarDist}), the EMD, and hence $t$-closeness, of the first equivalence class of Table \ref{TAB_IncidentsAnon} is
\begin{gather}
E'(\textbf{P}_{2,1},\textbf{Q}_2)
=\frac{1}{2}\left[\left|\frac{3}{3}-\frac{5}{14}\right| +
	\left|0-\frac{1}{14}\right| +
	\left|0-\frac{2}{14}\right| +\right.	
	\nonumber\\
\left.\left|0-\frac{1}{14}\right| +
	\left|0-\frac{3}{14}\right| +
	\left|0-\frac{1}{14}\right| +
	\left|0-\frac{1}{14}\right|\right]
	\nonumber\\
\approx 0.6429.
\end{gather}
Repeating this approach, the $t$-closeness values of the other classes are $E'(\textbf{P}_{2,2},\textbf{Q}_2)\approx 0.7143$, $E'(\textbf{P}_{2,3},\textbf{Q}_2)\approx 0.4286$, and $E(\textbf{P}_{2,4},\textbf{Q}_2)\approx 0.4429$, respectively. In the end, the $t$-closeness of Table \ref{TAB_IncidentsAnon} is the maximum of these four values, that is $0.7143$.

\section{Salary Example}
\label{SEC_ExSalary}

Consider the example put forward by Li et al., which is replicated in this paper in Table \ref{TAB_ExSalary} \cite{REF_tClose}. The quasi-identifiers \textit{Zip Code} and \textit{Age} are partially obfuscated so as to prevent one from precisely linking an individual in Table \ref{TAB_ExSalary} to the sensitive attributes \textit{Salary} and \textit{Disease}. This example makes use of the attribute \textit{Salary} and the nine associated values, organized into the set $Q'_3 =\{3\textrm{k}, 4\textrm{k}, 5\textrm{k}, 6\textrm{k}, 7\textrm{k}, 8\textrm{k}, 9\textrm{k}, 10\textrm{k}, 11\textrm{k}\}$, all $m = 9$ of which are distinct. Reiterating the point made in Section \ref{SUBSEC_NumAtt}, the values of $Q'_3$, although numbers, should be regarded as nothing more than labels. Each element $v_j\in Q'_3$ is unique, resulting in nine uniform holes of size $q_1 = q_2 = q_3 = \ldots = q_9 = 1/|Q'_3| =\sfrac{1}{9}$. Hence, $\textbf{Q}_3 =\left(\sfrac{1}{9},\sfrac{1}{9},\sfrac{1}{9}, \sfrac{1}{9},\sfrac{1}{9},\sfrac{1}{9},\sfrac{1}{9}, \sfrac{1}{9},\sfrac{1}{9}\right)$.
\begin{table}[h]
\centering
\begin{tabular}{rrrl}
\hline
	Zip Code & Age & Salary & Disease\\
\hline
	$476$** & $2$* & $3\textrm{k}$ & gastric ulcer\\
	$476$** & $2$* & $4\textrm{k}$ & gastritis\\
	$476$** & $2$* & $5\textrm{k}$ & stomach cancer\\
\hline
	$4790$* & $\geq 40$ & $6\textrm{k}$ & gastritis\\
	$4790$* & $\geq 40$ & $11\textrm{k}$ & flu\\
	$4790$* & $\geq 40$ & $8\textrm{k}$ & bronchitis\\
\hline
	$476$** & $3$* & $7\textrm{k}$ & bronchitis\\
	$476$** & $3$* & $9\textrm{k}$ & pneumonia\\
	$476$** & $3$* & $10\textrm{k}$ & stomach cancer\\
\hline
\end{tabular}
\caption{Anonymized salary data (based on Table 4 of Li et al. \cite{REF_tClose})}
\label{TAB_ExSalary}
\end{table}

The EMD is calculated for each of the three equivalence classes of Table \ref{TAB_ExSalary}. The first equivalence class, corresponding to the first three rows of Table \ref{TAB_ExSalary}, is given by $P'_{3,1} =\{3\textrm{k}, 4\textrm{k}, 5\textrm{k}\}\subseteq Q'_3$. The three elements $3\textrm{k}$, $4\textrm{k}$, and $5\textrm{k}$ of $P'_{3,1}$ each have a frequency of one, giving masses of $p_1 = p_2 = p_3 = 1/|P'_{3,1}| =\sfrac{1}{3}$. The six other values of $Q'_3$, not found in $P'_{3,1}$, have masses of $p_4 = p_5 = p_6 = p_7 = p_8 = p_9 = 0$. Consequently, $\textbf{P}_{3,1} =\left(\sfrac{1}{3},\sfrac{1}{3},\sfrac{1}{3}, 0, 0, 0, 0, 0, 0\right)$. The second equivalence class of Table \ref{TAB_ExSalary}, $P'_{3,2} =\{6\textrm{k}, 8\textrm{k}, 11\textrm{k}\}$, has a distribution of $\textbf{P}_{3,2} =\left(0, 0, 0,\sfrac{1}{3}, 0, \sfrac{1}{3}, 0, 0,\sfrac{1}{3}\right)$, and the final equivalence class of Table \ref{TAB_ExSalary}, namely $P'_{3,3} =\{7\textrm{k}, 9\textrm{k}, 10\textrm{k}\}$, corresponds to the distribution $\textbf{P}_{3,3} =\left(0, 0, 0, 0,\sfrac{1}{3}, 0,\sfrac{1}{3},\sfrac{1}{3}, 0\right)$.

\subsection{Calculation using the Definition}
\label{SUBSEC_SalaryCaseStudyDefn}

\begin{figure*}[h]
\centering
\includegraphics[]{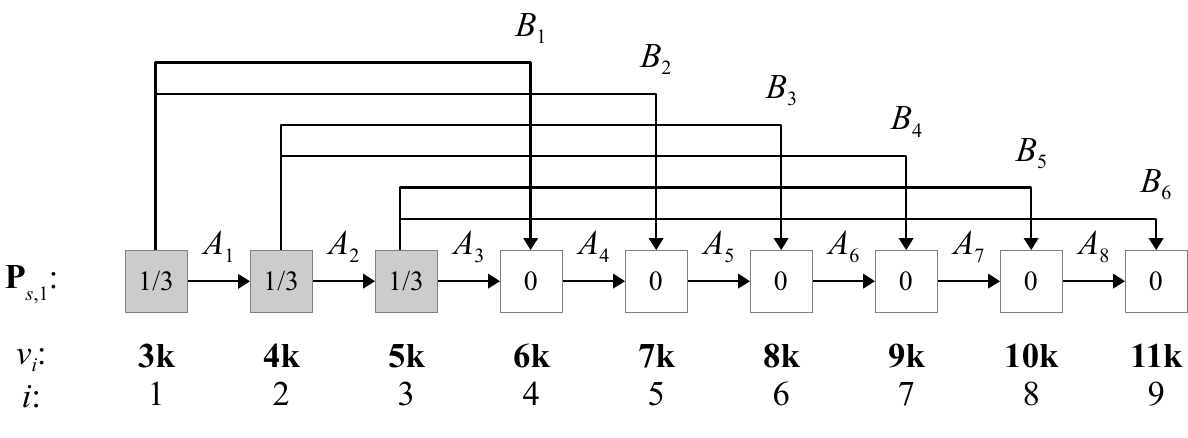}
\caption{Transformation of $\textbf{P}_{3,1}$ to $\textbf{Q}_3$}
\label{FIG_ExSalary}
\end{figure*}

The first task is to determine the cost of transforming $\textbf{P}_{3,1}$ into $\textbf{Q}_3$. Using Equation (\ref{EQ_EMD}), this is calculated as
\begin{gather}
E(\textbf{P}_{3,1},\textbf{Q}_3)
=\frac{1}{9-1}\left[
	\left|\frac{1}{3}-\frac{1}{9}\right| +\right.
	\nonumber\\
\left.\left|\left(\frac{1}{3}-\frac{1}{9}\right) +
	\left(\frac{1}{3}-\frac{1}{9}\right)\right| +\right.
	\nonumber\\
\left.\left|\left(\frac{1}{3}-\frac{1}{9}\right) +
	\left(\frac{1}{3}-\frac{1}{9}\right) +
	\left(\frac{1}{3}-\frac{1}{9}\right)\right|+\right.
	\nonumber\\
\left.\left|\left(\frac{1}{3}-\frac{1}{9}\right) +
	\left(\frac{1}{3}-\frac{1}{9}\right) +\right.\right.
	\nonumber\\
\left.\left.\left(\frac{1}{3}-\frac{1}{9}\right) +
	\left(0-\frac{1}{9}\right)\right| +
	\cdots\right]\label{EQ_ExSalaryDefnExpansion}\\
=\frac{1}{8}\left[
	\underbracket[0.5pt]{\frac{2}{9}}_{A_1} +
	\underbracket[0.5pt]{\frac{4}{9}}_{A_2} +
	\underbracket[0.5pt]{\frac{6}{9}}_{A_3} +
	\underbracket[0.5pt]{\frac{5}{9}}_{A_4} +
	\underbracket[0.5pt]{\frac{4}{9}}_{A_5} +
	\underbracket[0.5pt]{\frac{3}{9}}_{A_6} +
	\underbracket[0.5pt]{\frac{2}{9}}_{A_7} +
	\underbracket[0.5pt]{\frac{1}{9}}_{A_8}
	\right]\label{EQ_ExSalaryDefnSimplification}\\
= 0.375.
\label{EQ_ExSalaryDefnResult}
\end{gather}
The individual ``actions'' $A_1, A_2, A_3, \ldots, A_8$ of Equation (\ref{EQ_ExSalaryDefnSimplification}), shown in Figure \ref{FIG_ExSalary}, correspond to the movements of masses that transform $\textbf{P}_{3,1}$ into $\textbf{Q}_3$. In $A_1$, $\sfrac{2}{9}$ from mass $p_1 =\sfrac{1}{3}$ of $v_1 = 3\textrm{k}$ is moved forward to $v_2$. Then, in $A_2$, $\sfrac{2}{9}$ from mass $p_2 =\sfrac{1}{3}$ of $v_2 = 4\textrm{k}$, along with the $\sfrac{2}{9}$ from $v_1$, thus $\sfrac{2}{9} +\sfrac{2}{9} =\sfrac{4}{9}$, is moved forward to $v_3$. A further $\sfrac{2}{9}$ is picked up at $v_3$, yielding a total of $\sfrac{4}{9} +\sfrac{2}{9} =\sfrac{6}{9}$. The first hole is encountered at $v_4$, at which point $\sfrac{1}{9}$ is deposited, leaving
$\sfrac{6}{9} -\sfrac{1}{9} =\sfrac{5}{9}$ to carry forward. In each of the remaining five holes of $v_5$ through $v_9$, a mass of $\sfrac{1}{9}$ is dropped, thereby depleting all of the available mass. Following the actions of $A_1, A_2, A_3, \ldots, A_8$, each element of $\textbf{P}_{3,1}$ contains a mass of $\sfrac{1}{9}$, meaning that $\textbf{P}_{3,1}$ has been transformed into the (uniform) distribution $\textbf{Q}_3$.

In the case of the two remaining equivalence classes $P'_{3,2}$ and $P'_{3,3}$, $\textbf{P}_{3,2}$ and $\textbf{P}_{3,3}$ can be individually transformed, again using Equation (\ref{EQ_EMD}), to $\textbf{Q}_3$ at costs of (approximately) $0.1667$ and $0.2361$, respectively. Lastly, given the three EMD values above, the $t$-closeness of Table \ref{TAB_ExSalary} is determined to be $t=\max(0.375, 0.1667, 0.2361) = 0.375$.

\subsection{Calculation using the Weighted Ordered Distance}
\label{SUBSEC_ExSalaryWeightedOrderedDist}

The cost of changing $\textbf{P}_{3,1}$ into $\textbf{Q}_3$ can also be found using the weighted ordered distance approach of Equation (\ref{EQ_EMDWeightedOrderedDist}). The full process of doing so is given in the six steps $B_1$, $B_2$, $B_3$, \ldots, $B_6$ below.

\paragraph{Transformation of $\textbf{P}_{3,1}$ to $\textbf{Q}_3$:}
\begin{enumerate}[label=$B_\arabic*$.]
\item Move $w_{1,4}=\sfrac{1}{9}$ from
	$p_1 =\sfrac{1}{3}$ of
	$v_1 = 3\textrm{k}\in P'_{3,1}$
	from index $i=1$ to $j=4$
	($i, j\in\textbf{P}_{3,1}$), leaving a remainder of
	$p_1 =\sfrac{1}{3} -\sfrac{1}{9} =\sfrac{2}{9}$.
\item Move $w_{1,5}=\sfrac{1}{9}$ from
	the remainder of $p_1 =\sfrac{2}{9}$ of
	$v_1 = 3\textrm{k}\in P'_{3,1}$ from
	$i=1$ to $j=5$ ($i, j\in\textbf{P}_{3,1}$),
	leaving $p_1 =\sfrac{2}{9} -\sfrac{1}{9} =
	\sfrac{1}{9}$.
\item Move $w_{2,6}=\sfrac{1}{9}$ from
	$p_2 =\sfrac{1}{3}$ of $v_2$ from
	$i=2$ to $j=6$, leaving $p_2 =\sfrac{2}{9}$.
\item Move $w_{2,7}=\sfrac{1}{9}$ from
	$p_2 =\sfrac{2}{9}$ of $v_2$ from
	$i=2$ to $j=7$, leaving $p_2 =\sfrac{1}{9}$.
\item Move $w_{3,8}=\sfrac{1}{9}$ from
	$p_3 =\sfrac{1}{3}$ of $v_3$ from
	$i=3$ to $j=8$, leaving $p_3 =\sfrac{2}{9}$.
\item Move $w_{3,9}=\sfrac{1}{9}$ from
	$p_3 =\sfrac{2}{9}$ of $v_3$ from
	$i=3$ to $j=9$, leaving $p_3 =\sfrac{1}{9}$.
\end{enumerate}
Using Equation (\ref{EQ_WeightedOrderedDist}), the EMD, and thus $t$-closeness, of $\textbf{P}_{3,1}$ and $\textbf{Q}_3$ is calculated as
\begin{gather}
E(\textbf{P}_{3,1},\textbf{Q}_3)
=\underbracket[0.5pt]
	{\frac{1}{9}\cdot\frac{|1-4|}{9-1}}_{B_1} +	
	\underbracket[0.5pt]
	{\frac{1}{9}\cdot\frac{|1-5|}{9-1}}_{B_2} +	
	\underbracket[0.5pt]
	{\frac{1}{9}\cdot\frac{|2-6|}{9-1}}_{B_3} +
	\nonumber\\
\underbracket[0.5pt]	
	{\frac{1}{9}\cdot\frac{|2-7|}{9-1}}_{B_4} +
	\underbracket[0.5pt]
	{\frac{1}{9}\cdot\frac{|3-8|}{9-1}}_{B_5} +
	\underbracket[0.5pt]
	{\frac{1}{9}\cdot\frac{|3-9|}{9-1}}_{B_6}
= 0.375.
\label{EQ_ExSalaryWeightedOrderedDist1}
\end{gather}
The result is the same as that of Equation (\ref{EQ_ExSalaryDefnResult}). The actions $B_1$, $B_2$, $B_3$, \ldots, $B_6$ of this transformation are seen in Equation (\ref{EQ_ExSalaryWeightedOrderedDist1}) and depicted in Figure \ref{FIG_ExSalary}. Observe that two actions, namely $B_1$ and $B_2$, ``pass through'' the ``region'' above action $A_1$ of Figure \ref{FIG_ExSalary}. These two actions correspond to term $A_1$ ($\sfrac{2}{9}$) of Equation (\ref{EQ_ExSalaryDefnSimplification}). Likewise, the four actions $B_1$, $B_2$, $B_3$, and $B_4$ that ``pass through'' the ``region'' of $A_2$ match up to $A_2$ ($\sfrac{4}{9}$) of Equation (\ref{EQ_ExSalaryDefnSimplification}). The six actions $B_1$, $B_2$, $B_3$, $B_4$, $B_5$, and $B_6$ of the ``region'' of $A_3$ relate to term $A_3$ ($\sfrac{6}{9}$) of Equation (\ref{EQ_ExSalaryDefnSimplification}), and so forth. Each time that an action ``passes through'' a ``region'', it contributes a mass of $\sfrac{1}{9}$ to that ``region''. These individual masses of $\sfrac{1}{9}$ are then collectively summed over a given ``region'', just as in Equation (\ref{EQ_ExSalaryDefnSimplification}), thereby linking Equations (\ref{EQ_ExSalaryDefnSimplification}) and (\ref{EQ_ExSalaryWeightedOrderedDist1}). Ultimately, both interpretations of the EMD presented in Equations (\ref{EQ_EMD}) and (\ref{EQ_EMDWeightedOrderedDist}) are equivalent.

Taking the same approach as above, one obtains
\begin{gather}
E(\textbf{P}_{3,2},\textbf{Q}_3)
=\frac{1}{9}\cdot\frac{|4-1|}{9-1} +
	\frac{1}{9}\cdot\frac{|4-2|}{9-1} +
	\frac{1}{9}\cdot\frac{|4-3|}{9-1} +
	\nonumber\\
\frac{1}{9}\cdot\frac{|6-4|}{9-1} +
	\frac{1}{9}\cdot\frac{|6-5|}{9-1} +
	\frac{1}{9}\cdot\frac{|9-7|}{9-1} +
	\frac{1}{9}\cdot\frac{|9-8|}{9-1}
	\nonumber\\
\approx 0.1667.
\end{gather}

The EMD of the final equivalence class of Table \ref{TAB_ExSalary}, specifically $P'_{3,3}$, is
\begin{gather}
E(\textbf{P}_{3,3},\textbf{Q}_3)
=\frac{1}{9}\cdot\frac{|5-1|}{9-1} +
	\frac{1}{9}\cdot\frac{|5-2|}{9-1} +
	\frac{1}{9}\cdot\frac{|5-3|}{9-1} +
	\nonumber\\
\frac{1}{9}\cdot\frac{|7-4|}{9-1} +
	\frac{1}{9}\cdot\frac{|7-5|}{9-1} +
	\frac{1}{9}\cdot\frac{|7-6|}{9-1} +		
	\nonumber\\
\frac{1}{9}\cdot\frac{|8-7|}{9-1} +
	\frac{1}{9}\cdot\frac{|8-9|}{9-1}
\approx 0.2361.
\end{gather}

\section{Merit Points Example}
\label{SEC_ExMeritPts}

Observe that in all four distributions of Section \ref{SEC_ExSalary}, each of the elements of $Q'_3$, $P'_{3,1}$, $P'_{3,2}$, $P'_{3,3}$ has an individual frequency of one. The EMD also extends to sets that are strictly multisets (having repeated values). Let $Q'_4 =\{3, 3, 3, 3, 4, 4, 4, 1, 1, 2\}$ represent the sensitive values of the attribute \textit{Merit Points} of Table \ref{TAB_ExMeritPts}. With four values of $3$, three instances of $4$, two values of $1$, and a single $2$, of a total of $m = 10$ elements, the accompanying distribution is $\textbf{Q}_4 =\{\sfrac{4}{10},\sfrac{3}{10}, \sfrac{2}{10},\sfrac{1}{10}\}$. The four equivalence classes of $\textbf{Q}_4$ are $P'_{4,1} =\{4, 1, 2\}\subseteq Q'_4$ ($\textbf{P}_{4,1} =\{0, \sfrac{1}{3},\sfrac{1}{3}, \sfrac{1}{3}\}$), $P'_{4,2} =\{3\}$ ($\textbf{P}_{4,2} =\{1, 0, 0, 0\}$), $P'_{4,3} =\{3, 3, 4, 1\}$ ($\textbf{P}_{4,3} =\{\sfrac{2}{4}, \sfrac{1}{4},\sfrac{1}{4}, 0\}$), and $P'_{4,4} =\{3, 4\}$ ($\textbf{P}_{4,4} =\{\sfrac{1}{2},\sfrac{1}{2}, 0, 0\}$). Note that the explicit identifier \textit{Project Name} of Table \ref{TAB_ExMeritPts} shows only the first letter of each name.
\begin{table}[h]
\centering
\begin{tabular}{lr}
\hline
	Project Name & Merit Points\\
\hline
	$E$** & $1$\\
	$E$** & $4$\\
	$E$** & $2$\\
\hline
	$U$** & $3$\\
\hline
	$G$** & $3$\\
	$G$** & $4$\\
	$G$** & $3$\\
	$G$** & $1$\\
\hline
	$R$** & $4$\\
	$R$** & $3$\\
\hline
\end{tabular}
\caption{Anonymized merit points data}
\label{TAB_ExMeritPts}
\end{table}

\subsection{Calculation using the Definition}
\label{SUBSEC_MeritPtsCaseStudyDefn}

Applying Equation (\ref{EQ_EMD}) to distributions $\textbf{P}_{4,1}$ and $\textbf{Q}_4$ of the attribute \textit{Merit Points}, one obtains
\begin{gather}
E(\textbf{P}_{4,1},\textbf{Q}_4)
=\frac{1}{4-1}
	\left[\left|0-\frac{4}{10}\right| +\right.
	\nonumber\\
\left.\left|\left(0-\frac{4}{10}\right) +
	\left(\frac{1}{3}-\frac{3}{10}\right)\right| +\right.
	\nonumber\\
\left.\left|\left(0-\frac{4}{10}\right) +
	\left(\frac{1}{3}-\frac{3}{10}\right) +
	\left(\frac{1}{3}-\frac{2}{10}\right)\right|+\right.
	\nonumber\\
\left.\left|\left(0-\frac{4}{10}\right) +
	\left(\frac{1}{3}-\frac{3}{10}\right) +\right.\right.
	\nonumber\\
\left.\left.\left(\frac{1}{3}-\frac{2}{10}\right) +
	\left(\frac{1}{3}-\frac{1}{10}\right)\right|
	\right]\label{EQ_ExMeritPtsDefnExpansion}\\
\approx 0.3333.
\end{gather}

Similarly, $E(\textbf{P}_{4,2},\textbf{Q}_4)\approx 0.3333$,
$E(\textbf{P}_{4,3},\textbf{Q}_4) = 0.0833$, and $E(\textbf{P}_{4,4},\textbf{Q}_4)\approx 0.1667$. Combining the results of this example, the $t$-closeness of Table \ref{TAB_ExMeritPts} is equal to $t = 0.3333$, the maximum of the four EMD values.

\subsection{Weighted Ordered Distance}
\label{SUBSEC_MeritPtsCaseStudyWeightedOrderedDist}

The task of transferring the masses of $\textbf{P}_{4,1}$ to the holes of $\textbf{Q}_4$ of Table \ref{TAB_ExMeritPts}, thus transforming $\textbf{P}_{4,1}$ to $\textbf{Q}_4$, is given in the actions $C_1$, $C_2$, and $C_3$ that follow.

\paragraph{Transformation of $\textbf{P}_{4,1}$ to $\textbf{Q}_4$:}

\begin{enumerate}[label=$C_\arabic*$.]
\item Move $w_{2,1}=\sfrac{1}{30}$ from
	$p_2 =\sfrac{1}{3}$ of
	$v_2 = 4\in P'_{4,1}$ from
	index $i=2$ to $j=1$
	($i, j\in\textbf{P}_{4,1}$), leaving a remainder of
	$p_2 =\sfrac{1}{3} -\sfrac{1}{30} =\sfrac{3}{10}$.
\item Move $w_{3,1}=\sfrac{4}{30}$ from
	$p_3 =\sfrac{1}{3}$ of $v_3 = 1\in P'_{4,1}$  from
	$i=3$ to $j=1$ ($i, j\in\textbf{P}_{4,1}$),
	leaving $p_2 =\sfrac{1}{3} -\sfrac{4}{30} =\sfrac{2}{10}$.
\item Move $w_{4,1}=\sfrac{7}{30}$ from
	$p_4 =\sfrac{1}{3}$ of $v_4 = 2$ from 
	$i=4$ to $j=1$, leaving $p_4 =\sfrac{1}{10}$.
\end{enumerate}

Using Equation (\ref{EQ_EMDWeightedOrderedDist}), one finds that
\begin{gather}
E(\textbf{P}_{4,1},\textbf{Q}_4)
=\underbracket[0.5pt]
	{\frac{1}{30}\cdot\frac{|2-1|}{4-1}}_{C_1} +
	\underbracket[0.5pt]
	{\frac{4}{30}\cdot\frac{|3-1|}{4-1}}_{C_2} +
\underbracket[0.5pt]
	{\frac{7}{30}\cdot\frac{|4-1|}{4-1}}_{C_3}	
	\nonumber\\
\approx 0.3333.
\end{gather}

As before, $E(\textbf{P}_{4,2},\textbf{Q}_4)\approx 0.3333$,
$E(\textbf{P}_{4,3},\textbf{Q}_4) = 0.0833$, and $E(\textbf{P}_{4,4},\textbf{Q}_4)\approx 0.1667$.

\section{Disease Example}
\label{SEC_ExDisease}

As a final case, consider the sensitive attribute \textit{Disease} of Table \ref{TAB_ExSalary}. Globally, the values of \textit{Disease} are given as
\begin{gather*}
Q'_5 =\{\textrm{gastric~ulcer},
	\textrm{gastritis},
	\textrm{stomach~cancer},
	\textrm{gastritis},\\
	\textrm{flu},
	\textrm{bronchitis},
	\textrm{bronchitis},
	\textrm{pneumonia},
	\textrm{stomach~cancer}\}.
\end{gather*}
In the set $Q'_5$, there is one instance of ``gastric ulcer'', two of each of ``gastritis'' and ``stomach cancer'', one of ``flu'', two of ``bronchitis'', and finally one of ``pneumonia''. Accordingly, the distribution of these values over Table \ref{TAB_ExSalary} is $\textbf{Q}_5 =\{\sfrac{1}{9},\sfrac{2}{9},\sfrac{2}{9}, \sfrac{1}{9},\sfrac{2}{9},\sfrac{1}{9}\}$. In the first equivalence class of Table \ref{TAB_ExSalary}, in particular
\begin{gather*}
P'_{5,1} =\{\textrm{gastric~ulcer},
	\textrm{gastritis},
	\textrm{stomach~cancer}\},
\end{gather*}
there is one occurrence of each of ``gastric ulcer'', ``gastritis'', and ``stomach cancer'', yielding a distribution of $\textbf{P}_{5,1} =\{\sfrac{1}{3},\sfrac{1}{3},\sfrac{1}{3}, 0, 0, 0\}$. Given that \textit{Disease} is a categorical attribute, the EMD is calculated using Equation (\ref{EQ_EMDVarDist}) as follows
\begin{gather}
E'(\textbf{P}_{5,1},\textbf{Q}_5)
=\frac{1}{2}
\left[\left|\frac{1}{3}-\frac{1}{9}\right| +
	\left|\frac{1}{3}-\frac{2}{9}\right| +
	\left|\frac{1}{3}-\frac{2}{9}\right| +\right.
	\nonumber\\
\left.\left|0-\frac{1}{9}\right| +
	\left|0-\frac{2}{9}\right| +
	\left|0-\frac{1}{9}\right|\right]
\approx 0.4444.
\end{gather}

The second and third equivalence classes of the attribute \textit{Disease} of Table \ref{TAB_ExSalary} are
\begin{gather*}
P'_{5,2} =\{\textrm{gastritis},
	\textrm{flu},
	\textrm{bronchitis}\}
\end{gather*}
($\textbf{P}_{5,2} =\{0, \sfrac{1}{3}, 0, \sfrac{1}{3}, \sfrac{1}{3}, 0\}$)
and
\begin{gather*}
P'_{5,3} =\{\textrm{bronchitis},
	\textrm{pneumonia},
	\textrm{stomach~cancer}\}
\end{gather*}
($\textbf{P}_{5,3} =\{0, 0, \sfrac{1}{3}, 0, \sfrac{1}{3}, \sfrac{1}{3}\}$), respectively. The EMD of $\textbf{P}_{5,2}$ and $\textbf{P}_{5,3}$, also obtained via Equation (\ref{EQ_EMDVarDist}), are both (approximately) $0.4444$. Thus, the $t$-closeness of the attribute \textit{Disease} is $0.4444$.

\section{Efficient EMD Algorithm}
\label{SEC_EfficientEMDAlgorithm}

The redundancies of the calculation of the EMD via Equation (\ref{EQ_EMD}) are visible in the expansions of Equations (\ref{EQ_ExSalaryDefnExpansion}) and (\ref{EQ_ExMeritPtsDefnExpansion}). These expansions each contain repeated sequences of sums of differences of the form
\begin{gather}
E(\textbf{P},\textbf{Q})
=\frac{1}{m-1}
\left[\left|p_1 - q_1\right| +
	\left|\left(p_1 - q_1\right) +
	\left(p_2 - q_2\right)\right| +\right.
	\nonumber\\
\left.\left|\left(p_1 - q_1\right) +
	\left(p_2 - q_2\right) +
	\left(p_3 - q_3\right)\right| + \ldots\right],
\end{gather}
meaning that, for example, the sum $\left(p_1 - q_1\right) +\left(p_2 - q_2\right)$ is repeatedly calculated. Because of these redundancies, the na{\"{i}}ve computation of the EMD via Equation (\ref{EQ_EMD}) can be replaced by the enhanced procedure given in Algorithm \ref{ALG_EfficientEMD} \cite{REF_EMDNumEfficient}. Algorithm \ref{ALG_EfficientEMD} requires only a single pass over the data, giving a linear run-time complexity of $O(m)$, which is less than the quadratic complexity
$O(m\cdot (1 + 2 + 3 +\cdots + m)) = O(m^2)$
of the na{\"{i}}ve implementation of Equation (\ref{EQ_EMD}).
\begin{algorithm}[h]
\begin{algorithmic}
\State $\textrm{EMD}\gets 0$\Comment{initialize EMD}
\State $S\gets 0$\Comment{initialize (current) sum $S$}
\\
\For{$i=1~\textbf{to}~m$}
	\Comment{for each $p_i\in\textbf{P}$,
		$q_i\in\textbf{Q}$}
	\State $S\gets S +\left(p_i - q_i\right)$
		\Comment{increase $S$}
	\State $\textrm{EMD}\gets\textrm{EMD}+ |S|$
		\Comment{increase EMD}
\EndFor
\\
\State $\textrm{EMD}\gets\textrm{EMD}/(m-1)$\Comment{$m\neq 1$}
\end{algorithmic}
\caption{Efficient EMD}
\label{ALG_EfficientEMD}
\end{algorithm}

\section{Conclusion}
\label{SEC_Conclusion}

This paper presents three examples of the calculation of the one-dimensional EMD in the context of $t$-closeness. The first example examines the well-known scenario given by the designers of $t$-closeness, the second looks at more arbitrary distributions, and the third pertains to a categorical variable. Details not previously explained elsewhere in the literature are thoroughly articulated. The existing definition of the EMD is empirically demonstrated to be equivalent to the sums of the individual weighted ordered distances between masses and holes when optimal moves are made. As well, an efficient method of computing the EMD is presented.

\section*{Acknowledgements}

Funding for this research was provided by ISM Canada. Additional funding was provided by the Natural Sciences and Engineering Research Council of Canada (NSERC) through a Collaborative Research and Development Grant (CRDPJ 514906-17) and a Discovery Grant (RGPIN-2014-2014-04598) awarded to Hamilton. The authors wish to thank Rahim Samei of ISM Canada for his help in the review of this article.

\footnotesize
\bibliographystyle{unsrt}
\bibliography{tTutorial_arXiv}

\begin{thebibliography}{10}

\bibitem{REF_tClose}
Ninghui Li, Tiancheng Li, and Suresh Venkatasubramanian.
\newblock $t$-closeness: {P}rivacy beyond $k$-anonymity and $l$-diversity.
\newblock In {\em IEEE 23rd Int. Conf. Data Engineering}, 2007.

\bibitem{REF_tClose2}
Ninghui Li, Tiancheng Li, and Suresh Venkatasubramanian.
\newblock Closeness: {A} new privacy measure for data publishing.
\newblock {\em IEEE Trans. Knowledge and Data Engineering}, 22(7), 2010.

\bibitem{REF_LDiv}
Ashwin Machanavajjhala, Daniel Kifer, Johannes Gehrke, and Muthuramakrishnan
  Venkitasubramaniam.
\newblock $l$-diversity: {P}rivacy beyond $k$-anonymity.
\newblock {\em ACM Trans. Knowledge Discovery from Data}, 1(1), 2007.

\bibitem{REF_DBAnon}
Josep Domingo-Ferrer, David S{\'{a}}nchez, and Jordi Soria-Comas.
\newblock {\em Database Anonymization: {P}rivacy Models, Data Utility, and
  Microaggregation-based Inter-model Connections}.
\newblock Morgan \& Claypool, 2016.

\bibitem{REF_PPDP}
Benjamin~C.M. Fung, Ke~Wang, Ada Wai-Chee Fu, and Philip~S. Yu.
\newblock {\em Introduction to Privacy-Preserving Data Publishing: {C}oncepts
  and Techniques}.
\newblock CRC Press, 2010.

\bibitem{REF_EqClass}
Shafiullah Khan, Al-Sakib Pathan, and Nabil Alrajeh, editors.
\newblock {\em Wireless Sensor Networks: {C}urrent Status and Future Trends}.
\newblock Taylor \& Francis Group, 2012.

\bibitem{REF_kAnon}
Latanya Sweeney.
\newblock $k$-anonymity: a model for protecting privacy.
\newblock {\em Int. J. Uncertainty, Fuzziness and Knowledge-based Systems},
  10(5):557--570, 2002.

\bibitem{REF_CattClose}
Raymond Wong and Ada Fu.
\newblock {\em Privacy-Preserving Data Publishing: {A}n Overview}.
\newblock Morgan \& Claypool, 2010.

\bibitem{REF_FuzzyPartitions}
Pelayo Quir{\'{o}}s, Pedro Alonso, Irene D{\'{i}}az, and Susana Montes.
\newblock On the use of fuzzy partitions to protect data.
\newblock {\em Integrated Computer-Aided Engineering}, 21(4):355--366, 2014.

\bibitem{REF_KLDivSemanticPrivacy}
Chaofeng Sha, Yi~Li, and Aoying Zhou.
\newblock On $t$-{C}loseness with {KL}-divergence and semantic privacy.
\newblock In {\em Proc. 15th Int. Conf. Database Systems for Advanced
  Applications}, volume~II, pages 153--167, 2010.

\bibitem{REF_Health}
Lengdong Wu, Hua He, and Osmar Za{\"{i}}ane.
\newblock Utility of privacy preservation for health data publishing.
\newblock In {\em 2013 IEEE 26th Int. Symp. Computer-Based Medical Systems},
  2013.

\bibitem{REF_Microaggregation}
Jordi Soria-Comas, Josep Domingo-Ferrer, David S{\'{a}}nchez, and Sergio
  Mart{\'{i}}nez.
\newblock t-closeness through microaggregation: {S}trict privacy with enhanced
  utility preservation.
\newblock {\em IEEE Trans. Knowledge \& Data Engineering}, 27(11):3098--3110,
  2015.

\bibitem{REF_SABRE}
Jianneng Cao, Panagiotis Karras, Panos Kalnis, and Kian-Lee Tan.
\newblock {SABRE}: {A} {S}ensitive {A}ttribute {B}ucketization and
  {RE}distribution framework for $t$-closeness.
\newblock {\em The VLDB Journal}, 20(1):59--81, 2011.

\bibitem{REF_SurveyThesis}
Debaditya Roy.
\newblock Determining t in t-closeness using multiple sensitive attributes.
\newblock Master's thesis, National Institute of Technology {R}ourkela, 2013.

\bibitem{REF_tComplexity}
Hongyu Liang and Hao Yuan.
\newblock On the complexity of $t$-closeness anonymization and related
  problems.
\newblock In {\em Int. Conf. Database Systems for Advanced Applications}, pages
  331--345, 2013.

\bibitem{REF_EMD}
Yossi Rubner, Carlo Tomasi, and Leonidas Guibas.
\newblock The earth mover's distance as a metric for image retrieval.
\newblock {\em Int. J. Computer Vision}, 40(2):99--121, 2000.

\bibitem{REF_EMDFaceRecognition}
Enrico Vezzetti and Federica Marcolin.
\newblock {\em Similarity Measures for Face Recognition}.
\newblock Bentham Science Publishers, 2015.

\bibitem{REF_TotalOrder}
Daniel Velleman.
\newblock {\em How to Prove It: {A} Structured Approach}.
\newblock Cambridge University Press, 2nd edition, 2006.

\bibitem{REF_Multiset}
Gordon Pace.
\newblock {\em Mathematics of Discrete Structures for Computer Science}.
\newblock Springer-Verlag, 2012.

\bibitem{REF_ReflexiveAntisymmetricTransitive}
W.~Wallis.
\newblock {\em A Beginner's Guide to Discrete Mathematics}.
\newblock Birkh{\"a}user, 2003.

\bibitem{REF_Trichotomy}
Susanna Epp.
\newblock {\em Discrete Mathematics with Applications}.
\newblock Brooks/Cole, 4th edition, 2011.

\bibitem{REF_ManhattanDist}
Shashi Shekhar and Hui Xiong, editors.
\newblock {\em Encyclopedia of {GIS}}.
\newblock Springer, 2008.

\bibitem{REF_EMDNumEfficient}
Sung-Hyuk Cha and Sargur~N. Srihari.
\newblock On measuring the distance between histograms.
\newblock {\em Pattern Recognition}, 35(6):1355--1370, 2002.

\end{thebibliography}
\normalsize

\end{document}